\documentstyle[11pt]{article}

\title{\bf Pomeron fan diagrams with an infrared cutoff and
running coupling}
\author{M.A.Braun \thanks{Permanent address: 
Dep. High-Energy physics,
S.Petersburg University, 198504 S.Petersburg, Russia}\\
NCCU, Durham NC, USA}
\date{}
\def\beq{\begin{equation}}
\def\eeq{\end{equation}}

\begin{document}
\maketitle
\medskip
\vspace{1 cm}

{\bf Abstract}

By direct numerical calcultions the influence of a physically relevant
infrared cutoff and running coupling on the gluon density and
structure function of a large nucleus is studied in the perturbative
QCD approach. It is found that the infrared cutoff changes the
solutions very little. Running of the coupling produces a bigger
change, considerably lowering both the staruration momentum and values
of the structure functions.

\section{Introduction}
In the high-colour perturbative QCD  interaction of a probe with the nucleus
at low $x$ is described by a non-linear BFKL evolution equation,
which sums  pomeron fan diagrams
~\cite{bal,kov,bra}.
There has been
considerable activity to study the resulting nuclear structure functions and
gluon distributions ~\cite{AB,LT,LL,KKM,lub}. However one has to
remember that the 
BFKL dynamics put at the basis of this equation is only an approximation,
whose validity is restricted to not too small values of transverse momenta,
where the mere notion of gluons  becomes meaningless. Also the BFKL
dynamics does not take into account the running of the strong coupling constant.
In view of these limitations inherent in the perturbative approach to
low $x$ dynamics we consider it fruitless to study properties of the
non-linear evolution
equation as it stands at very low momenta (or large spatial
distances). In particular its generalization to include pomeron
dimensions much greater than those
of the target nucleus and derive consequences as to the behaviour of the
resulting amplitudes in the limit $x\to 0$ ~\cite{KW} are certainly
interesting from the mathematical point of view  but have little
physical relevance in our opinion.
Rather one has to study the dependence of the equation and its solutions 
on the infrared region and running of the coupling. Should such dependence
be very strong, the results obtained in the current approach, where
the coupling is fixed and no infrared cutoff is introduced, would have
little physical sense.

The first numerical solutions of the non-linear evolution equation have
shown that the resulting gluon density is concentrated at quite high
values of momenta,  around the so-called saturation momentum $Q_s(Y)$
which grows exponentially with the rapidity $Y=\log (1/x)$. This gives some
hope that the non-linear equation, unlike the linear one, is not sensitive
to the infrared region at all and in this way retains a full physical meaning in the
realistic world with confinement. However this point has not been fully
proven, due to the fact that the equation itself does not involve the
gluon density itself but some integral of it,  which is not at all
negligible in the infrared region and, in fact, grows logarithmically towards
small momenta. In this paper we intend to study the infrared dependence
of the non-linear evolution equation by direct numerical calculations.
Our results confirm that the solutions at small enough $x$ indeed  depend 
on the infrared region  only weakly. 

Another point
which we study is  inclusion of the running coupling.  Clearly this
cannot be done in any rigorous way.  We employ a simple intuitive model
for the running of the coupling, taking it dependent on the smallest
momentum in a given 3-gluon vertex.  Our calculations show that with the
running coupling the solutions 
do not change qualitatively, but the quantitative change is quite noticeable.
In particular the slope of the dependence of the saturation momentum on
$Y$ drops by a factor 2$\div$3, so that its values go down by 4 orders 
of magnitude at largest $Y$ studied.
The resulting structure functions also drop by an order of magnitude.  
We consider these results quite promising, since the  values of
$Q_s$ obtained from the original non-linear equation, without infrared
cutoff nor running of the coupling, are very large and grow unreasonably fast
with $Y$.

\section{The non-linear evolution equation with an infrared cutoff and
running coupling}
The non-linear evolution equation derived in ~\cite{kov,bra} for an
extended target reads
\beq
\Big(\frac{\partial}{\partial y}+H\Big)\phi(y,k)=-\phi^2(y,k).
\eeq
Here  $y=(\alpha_sN_c/\pi)Y$, $H$ is the BFKL Hamiltonian and
$\phi(y,k)$ is a Fourier transform of $\Phi(Y,r)/(2\pi r^2)$ where 
$\Phi$ has a meaning of the cross-section for the scattering of a colour dipole
on a target at a given impact parameter $b$. In fact both
$\phi$ and $\Phi$ also depend on $b$ through the initial condition at $y=0$.
 Eq. (1) is infrared stable, that is, preserves its meaning
when $k$ varies over the whole positive axis. Of course in numerical
calculations one has to limit these values at both small and large $k$.
Typically in our calculations ~\cite{bra,AB} we chose $k_{min}\sim
1.E-15$ GeV/c and
$k_{max}\sim 1.E+40$ GeV/c. With these values  the solution does not 
change when  the interval of $k$ is taken still larger.
Obviously these cutoffs served to a  purely
calculational purpose and the obtained solutions in fact correspond to the
completely uncut equation.

The physical infrared cutoff has to be of the order $\Lambda_{QCD}\sim 0.3$
GeV/c.
We may introduce it into the equation in two different ways. One is simply to
cut the allowed values of $k$ to $k>k_{min}$ and choose $k_{min}$ to be
around  $\Lambda_{QCD}$ (a "hard cutoff'" choice).  With such a cutoff the 
momenta of the intermediate real gluons are not cut and may be arbitrary small.
To also cut these latter, one may introduce an effective gluon mass of the same
order in all  gluon propagators and leave the overall cutoffs on momenta
the same as in the original equation (a "soft cutoff choice"). As we shall see,
our numerical calculations show that the resulting solutions are rather 
similar for both choices.

Passing to the running coupling, we recall that this is an unsolved problem
even for the linear BFKL equation. To see the
qualitative features of the solution, all one can do is to introduce the 
running coupling in a purely intuitive way, based on  scale arguments.
The 3-gluon BFKL vertex inside the BFKL Hamiltonian depends on 
the three gluon momenta $k_1,k_2,k_3$,
two of the virtual gluons and the third of the emitted real gluon. We choose to
introduce the running coupling for the vertex at a scale which is the
smallest of the  squares of these three  momenta 
$k_0^2=\min\{k_1^2,k_2^2,k_3^2\}$.  Using the soft cutoff
we have further to define the coupling for values of momenta below
$\Lambda _{QCD}$.  Our choice is to freeze the coupling below some
scale, for which we take the same scale $\Lambda_{QCD}$.
In our numerical calulations we have taken the frozen value of
the coupling constant $\alpha_s^{(0)}=0.2$.

\section{Numerical results}
\subsection{Gluon distribution}
We define the gluon distribution as in ~\cite{bra}
\beq
\frac{d xG(x,k^2)}{d^2bdk^2}=\frac{N_c}{2\pi^2\alpha_s}h(y,k),\ \
h(y,k)=k^2\nabla_k^2\phi(y,k).
\eeq
Our aim is to study the influence of the cutoff and running coupling on
this distribution.   For this aim the dependence on the impact parameter 
 is irrelevant, so that we shall assume
$\phi(y=0,k)$ and consequently $\phi(y,k^2)$ independent of $b$.
Physically
it corresponds to assuming a constant  nuclear profile function (approximation
of a "cylidrical nucleus").  For the initial distribution, following
~\cite{AB}, we chose the Golec-Biernat-Wuesthoff 
~\cite{GBW} form
\beq
\phi(0,k)=-\frac{1}{2}{\rm Ei}\,\left(-\frac{k^2}{0.3657}\right)
\eeq
where $k^2$ is in (GeV/c)$^2$.
In our calculations we compare four cases: no cutoff, no running coupling (case A),
hard cutoff, no running coupling (case B), soft cutoff, no running coupling
(case C) and finally soft cutoff, running coupling (case D). In all cases
the gluon distribution turns out to have a sharp maximum at a certain 
"saturation 
momentum" $Q_s(y)$, which grows  with $y$. With a fixed coupling (cases
A-C) to a good precision $Q_s(y)\propto \exp(\Delta y)$.
The slope $\Delta$ results practically independent of the introduced infrared
cutoff and its value lies between 2.2 and 2.3. However with a running
coupling  $Q_s(y)$ grows with $y$ much slowlier and not as the 
exponential, the slope
$\Delta$ diminishing from 1.0 at $y=5$ to 0.63 at $y=10$.
As a result,  values of the saturation momentum with
a running coupling are much lower than with a fixed one.
In Fig. 1 we show the saturation momentum $Q_s(y)$ for the described four
cases. With the running coupling the scaled rapidity $y$ has been defined 
as $y=(\alpha_s^{(0)}N_c/\pi)Y$, which implies that for comparison the fixed
coupling
has been taken equal to 0.2 for cases (A-C).
The form of the gluon distribution is shown 
in Figs. 2 and 3
where $h(y,k)$ is  plotted against the scaling variable $z=k/Q_s(y)$.
In Fig. 2 the distributions are presented for a relatively small
rapidity $y=3$, when the initial conditions are not completely forgotten
yet. One then observes a small difference in the three curves for fixed
coupling (A-C) especially noticeable at small values of $z$. The running
coupling curve is considerably different: it is narrower and its peak is
larger. This difference becomes still more pronounced at higher
rapidities, which is illustrated in Fig. 3, where we show distributions
for all four cases at $y=6$ and $y=10$ simultaneously. All fixed coupling
curves practically coincide for both rapidities, showing a clear
scaling behaviour, which has been discovered earlier for the original
solution without cutoffs ~\cite{bra,AB,lub}. So at these $y$ the
influence of the
physically relevant infrared cutoff is totally forgotten:
the gluon density does not change and remains scale invariant in spite of
the introduction of a scale of the order 0.3 GeV/c. This is exactly what was
conjectured in ~\cite{bra} :
at not too small $y$ the internal scale $Q_s$ generated by the
non-linear dynamics is much larger than infrared cutoffs, be it 0.3 GeV/c or
much smaller. 

On other hand, with the running coupling, the form of the gluon
distribution changes considerably.
Peaks of the running coupling curves become
nearly twice larger than for a fixed coupling and the curves themselve
become much  narrower. However the difference between the 
running coupling curves at $y=6$ and 10 is quite small,which indicates
that even with a running coupling, to a good approximation, scaling in $z$
is still observed.

\subsection{Structure function}
To see more physically noticeable consequences of the cutoffs and
running coupling we also calculated the structure function of Pb for the
mentioned 4 cases. For the initial function (now depending on $b$) we
have chosen the same eikonalized Golec-Biernat-Wuesthoff distribution
which was used in our earlier calculations according to the original
equation, so that we could read the results for case A directly from
~\cite{AB}.
Our results  are presented in Figs. 4 and 5 for $Q^2=100$ and 10000
(GeV/c)$^2$ respectively. The change due to the physical infrared cutoff
is now more pronounced: the structure functions with such a cutoff are
somewhat smaller than without cutoff, the difference growing with
$Q^2$. At $Q^2=10000$ (GeV/c)$^2$ introduction of a infrared cutoff lowers
the structure function  by $\sim$2 times at low $x$. Still the
behaviour in $x$ remains practically unchanged.
A more significant change occurs with a running coupling. The structure
function then grows with $1/x$ considerably slowlier. Its values at
$Q^2=10000$ (GeV/c)$^2$ and small $x$ become $\sim$4 times
smaller than without cutoffs and running coupling and this difference
seems to be growing at still smaller $x$.

\section{Conclusions}
By direct numerical calculations we studied the gluon density and
structure function of a large nucleus at small $x$ which follow
from the non-linear evolution equation with a physically 
reasonable infrared cutoff and also with a running coupling. 
Our results show that the gluon density does not change qualitatively.
In all cases it has  a strong peak at a certain saturation momentum,
$Q_s(y)$, which grows with $\log(1/x)$. The introduction of an
infrared cutoff of the order 0.3 GeV/c by itself does not practically
change the value of $Q_s$ nor the form of the gluon distribution
around it. Running of the coupling, on the other hand, does change
both: $Q_s(y)$ grows with $\log(1/x)$ much slowlier and not as an
exponential, the gluon distribution becomes narrower
and its height greater. Still in all cases scaling of the
distribution in $k/Q_s$ is preserved.

The structure functions go down with the introduction of an infrared
cutoff and especially with a running coupling. In the latter case
the growth of the structure function with $1/x$ is found to be 
considerably slower.

Our results confirm that the non-linear evolution equation is more or
less infrared stable, in contrast to the linear BFKL equation.
Changes introduced by an infrared cutoff are of no qualitative nature
and of minor quantitative influence. Running of the coupling
produces a somewhat bigger change. It is important that this change is in
the right direction. Fixed coupling  solutions lead to a
very fast growth of $Q_s(y)$ with $y$ and as a 
consequence to unreasonably large values for it. Introduction of the
running coupling considerably improves the situation.

\section{Acknowledgment}
The author is most thankful to B.Vlahovic for his interest in
this work and to NCCU, Durham, NC, USA for hospitality and financial
support.

\section{Figure captions}
Fig. 1. The saturation momentum $Q_s(y)$ as a function of
$y=(\alpha_sN_c/\pi)\ln(1/x)$. Curves from top to bottom
correspond to cases A,C,B and D(see the text). 
The lower curve corresponds to the running
coupling (case D). 

Fig. 2. The gluon distribution as a function of the scaling variable
$k/Q_s(y)$ at $y=3$. Curves with maxima from top to bottom
correspond to cases D,A,C and B (the first referring to the running
coupling)

Fig. 3 Same as in Fig. 3 for $y=6$ and 10. The two upper curves
correspond to the running coupling (the uppermost for $y=6$).
All the rest correspond to fixed coupling, either with an
infrared cutoff or without it.

Fig. 4. The structure function of Pb at $Q^2=100$ (GeV/c)$^2$.
Curves from top to bottom on the left correspond to cases
A, C, B and D (the last for the running coupling).

Fig. 5. Same as Fig. 4 for $Q^2=10000$ (GeV/c)$^2$ 
 
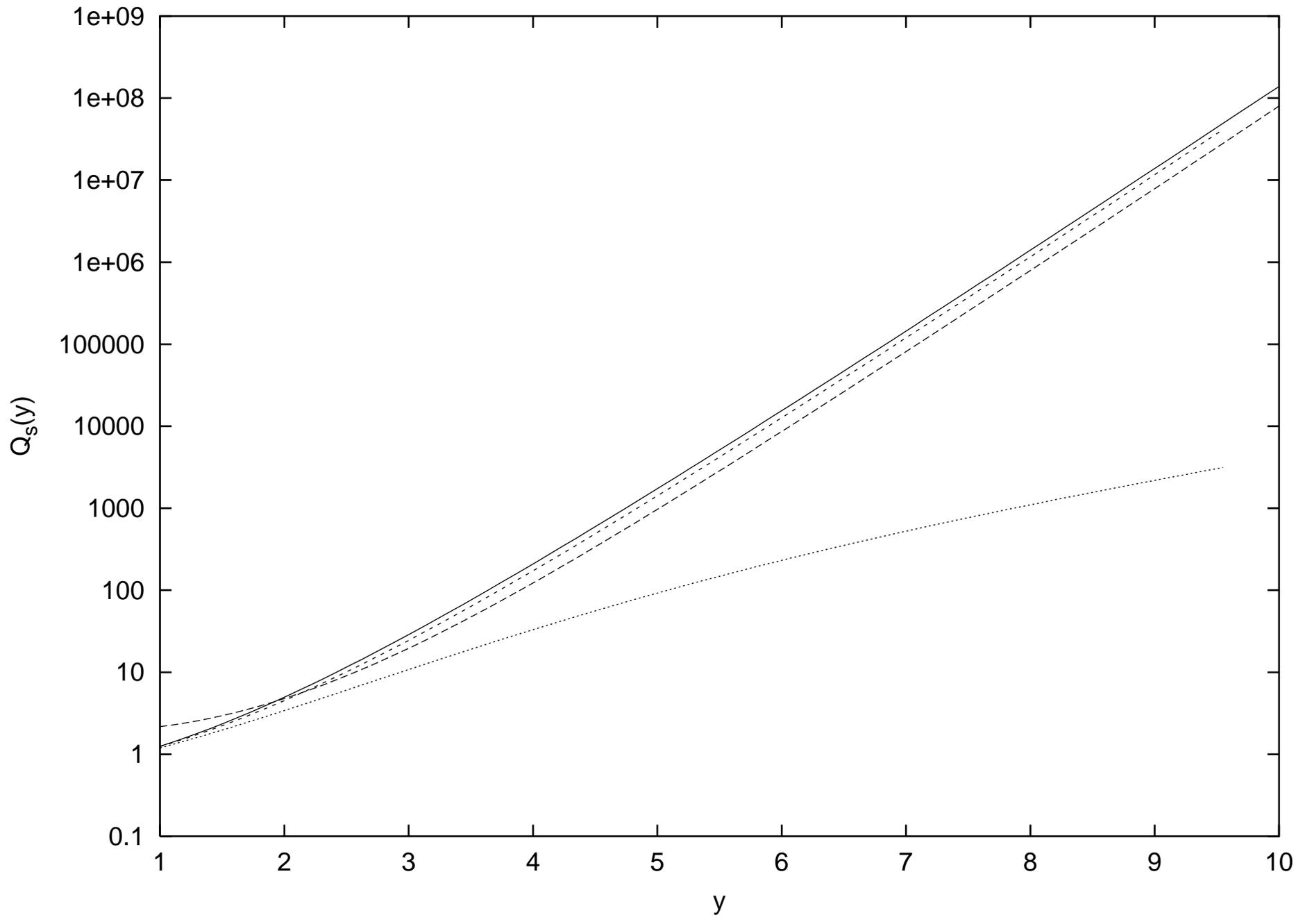
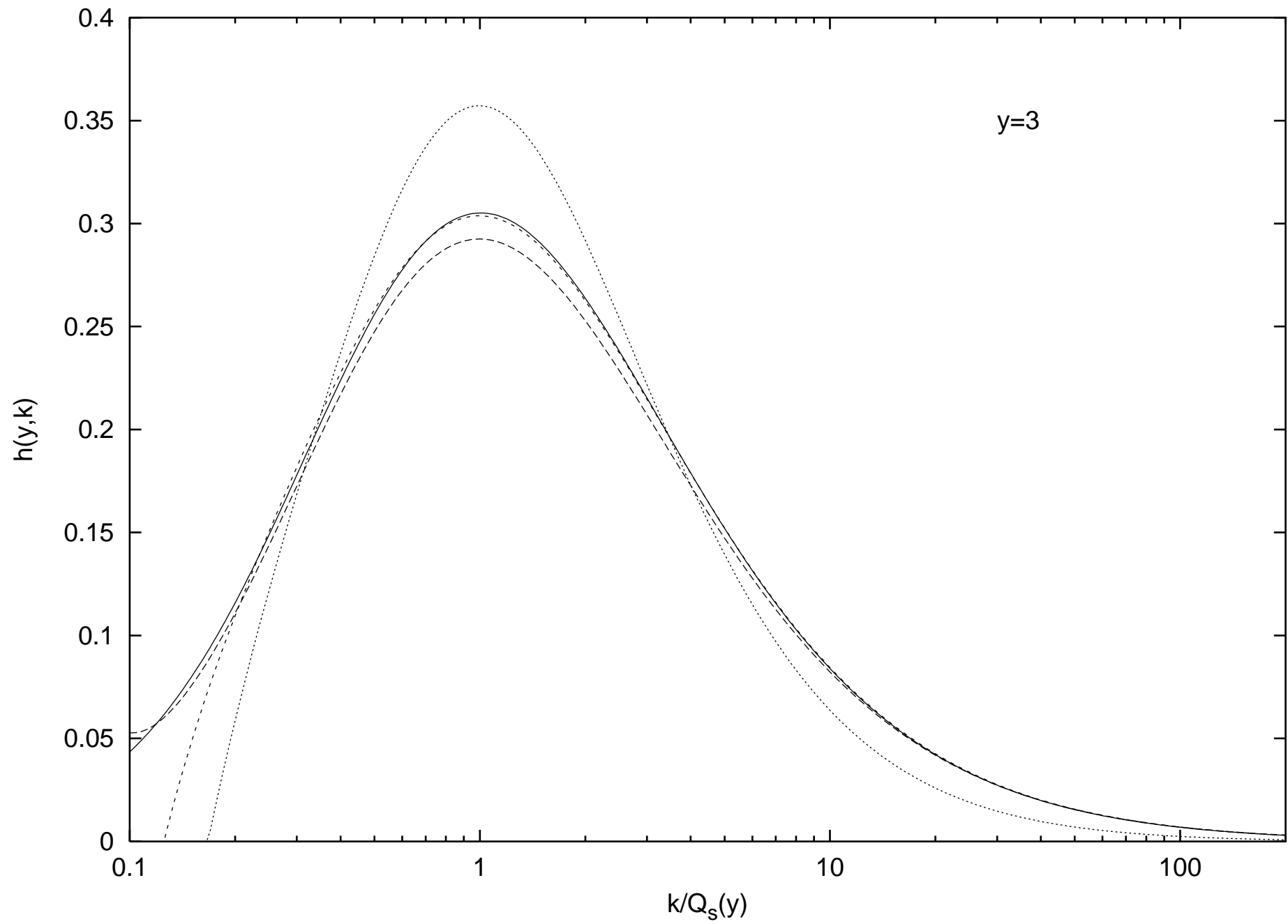
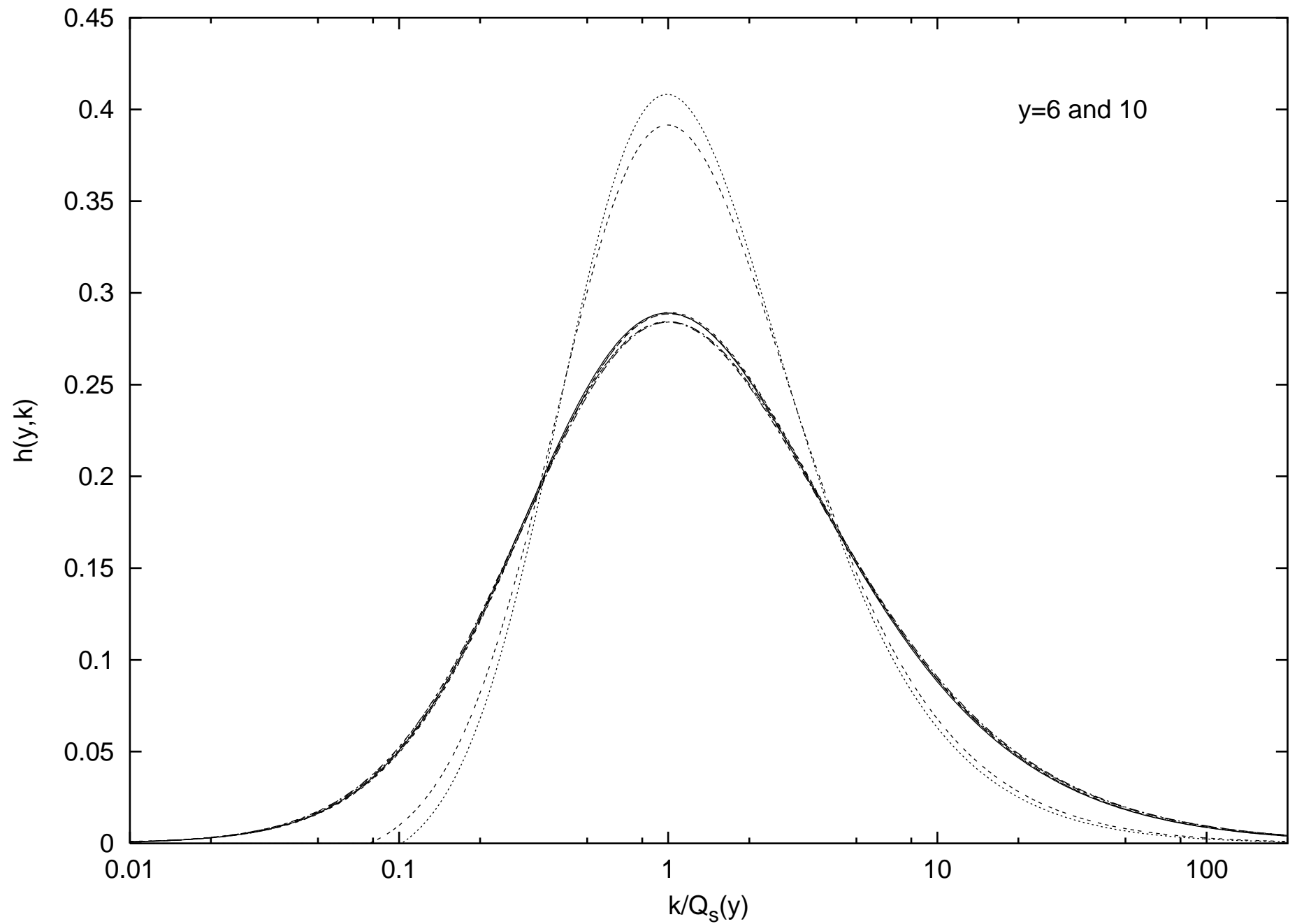
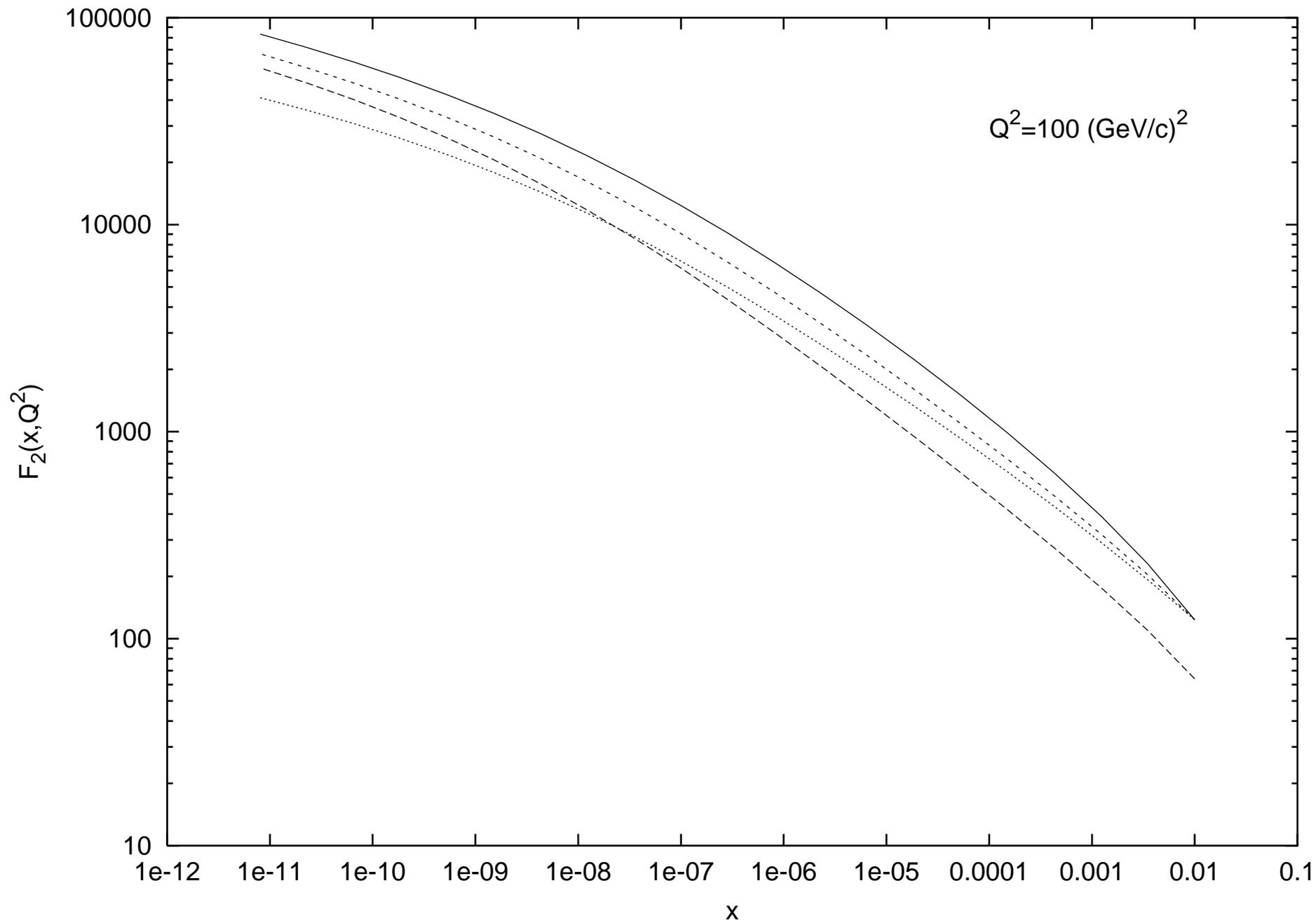
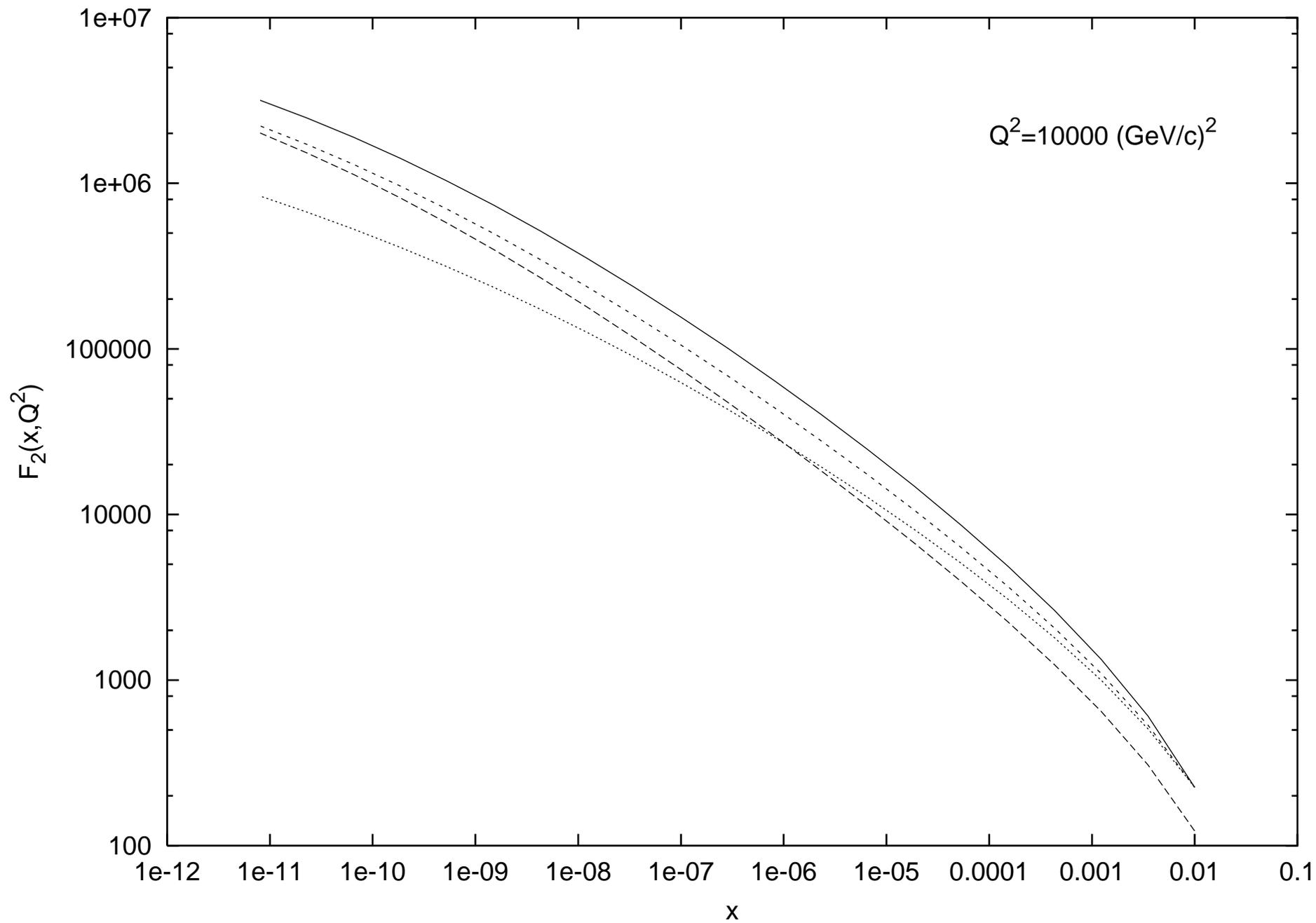
\end{document}